\documentstyle[12pt]{article}
\begin{document}

\thispagestyle{empty} \centerline{\large\bf An "Accidental"
Symmetry  Operator for the Dirac  Equation} \centerline{\large\bf
in the Coulomb Potential}
\bigskip
\begin{center}
Tamari T. Khachidze and Anzor A. Khelashvili\footnote{Email:
anzorkhelashvili@hotmail.com}\\ \vspace*{.5cm} {\em 3 Chavchavadze
Avenue, Department of Theoretical Physics,
\\Iv. Javakhishvili
Tbilisi State University, Tbilisi 0128, Georgia}\\ \vspace*{.5cm}
\end{center}
\vskip.5cm

\begin{abstract}
On the basis of the generalization of the theorem about K-odd
operators (K is the Dirac's operator), certain linear combination
is constructed, which appears to commute with the Dirac
Hamiltonian for Coulomb field. This operator coincides with the
Johnson and Lippmann operator and is intimately connected to the
familiar Laplace-Runge-Lenz  vector. Our approach guarantees not
only derivation of Johnson-Lippmann operator, but simultaneously
commutativity with the Dirac Hamiltonian follows.
\end{abstract}
\vspace*{.3cm}

\vspace*{1.0cm} PACS :  \\

\vspace*{0.3cm} Key words: :  Dirac's operator, hidden symmetry,
Hydrogen atom,  commutativity, supercharge, Witten superagebra.
\\

\newpage

Recently H.Katsura and H.Aoki \cite{KA} considered an exact
supersymmetry in the relativistic hydrogen atom in general
dimensions. It was established, that supercharges are connected to
the pseudoscalar invariant, which commutes with the Dirac
Hamiltonian in Coulomb field. In usual 3-dimensions this invariant
coincides to the Johnson-Lippmann (J.-L.) operator, which was
introduced by these authors in 1950 in a very brief abstract
\cite{JL}. As to more detailed settle, by our knowledge, it had
not been published neither then nor after.  First respond on
relativistic Kepler problem in this point of view appeared in
60-ies \cite{Bi}. Main attention was paid to the ways of deriving
simpler solutions of the Dirac equation as far as possible.

After 80-ies the interest was grown from the positions of
supersymmetric quantum mechanics. Supersymmetry of the Dirac
equation in Coulomb problem was demonstrated, but for this purpose
mainly the radial equations was considered for deriving the best
separation of variables.

In the above mentioned paper \cite{KA} the supercharges in
relativistic case is obtained for the first time. As it turns out,
the decisive part plays the J.-L. operator, the generalization of
which in arbitrary dimensions was introduced by these authors and,
therefore they named it as Johnson-Lippman -Katsura-Aoki operator.

In any article \cite{KA} or text books (See, \cite{Ber} , for
example), there is underlined only that its commutativity with
Hamiltonian can be proved by longtime and tedious calculations.

Derivation of Witten superalgebra is based on very simple logic:
It is known that the so called Dirac operator, which has the form
\cite{Ber}

\begin{equation}\label{Dirope}K=\beta\left(\vec{\Sigma}\vec{l}+1\right)~,\end{equation}
commutes with the Dirac Hamiltonian
\begin{equation}\label{Dirham}H=\vec{\alpha}\vec{p}+\beta m-{a\over r}~,~~~a\equiv Ze^2=Z\alpha~.\end{equation}
(In fact, $K$ commutes with Hamiltonian for arbitrary central
potential). In equation (\ref{Dirope}) $\vec{l}$ is the angular
momentum vector and $\vec{\Sigma}$ is the electron spin matrix
\[\vec{\Sigma}=\rho_1\vec{\alpha}=\gamma^5\vec{\alpha}=
\left(\begin{array}{cc}\vec{\sigma} & 0\\0 &
\vec{\sigma}\end{array}\right)~.\] The Sommerfeld formula for the
Hydrogen atom spectrum reads \cite{Dir}

\begin{equation}{E\over m}=\left[1+{\left(Z\alpha\right)^2\over
\left(n-|k|+\sqrt{k^2-\left(Z\alpha\right)^2}\right)^2}\right]^{-1/2}~.\end{equation}
Here $k$ is the eigenvalue of $K$, $|k|=\sqrt{j(j+1)+1/4}=j+1/2$.
It seems, that the energy spectrum is degenerate with respect of
signs of $k$, $\pm\mbox{sgn}(k)$.

An operator, that anticommute with $K$, is able to interchange
these signs, moreover if this operator would commute with
Hamiltonian at the same time, the supersymmetry algebra appears.
It was shown in \cite{KA}, that the J.-L. operator has precisely
these properties. The J.-L. operator looks like \cite{JL}
\begin{equation}\label{JLoper} A=\gamma^5\left\{{\vec{\alpha}\vec{r}\over r}-{i\over ma}K\gamma^5\left(H-\beta m\right)\right\}~.\end{equation}
We are inclined to think that in following importance of this
operator will enhance in studies of dynamical symmetries because
of its connection to the Laplace-Runge-Lenz (LRL) vector and
"accidental" degeneracy.

Therefore it seems highly desirable , by our opinion, to derive
the J.-L. operator from the first and simplest principles. Below
we propose one of our path  of construction this operator, which
is rather simple and very transparent.

First of all we need to extend one important theorem known for
Pauli electron \cite{AJ, BL} to the Dirac equation.

{\bf Theorem}: Suppose $\vec{V}$ is a vector with respect to the
orbital angular momentum $\vec{l}$, i.e.
\[\vec{l}\times\vec{V}+\vec{V}\times\vec{l}=2i\vec{V}~.\]
At the same time $\vec{V}$ is to be perpendicular to $\vec{l}$,
i.e., $\vec{l}\vec{V}=\vec{V}\vec{l}=0$.

Then the Dirac's $K$ operator anticommutes with $\vec{J}$- scalar
operator $\left(\vec{\Sigma}\vec{V}\right)$, i.e.,
\begin{equation}\left\{K,\left(\vec{\Sigma}\vec{V}\right)\right\}=0~.\end{equation}
The proof of this theorem is very easy. For that it is sufficient
to consider the product
$\left(\vec{\Sigma}\,\vec{l}+1\right)\left(\vec{\Sigma}\,\vec{V}\right)$
in direct and reversed orders, making use of commutating
properties of Dirac matrices.   It follows that
\[\left(\vec{\Sigma}\,\vec{l}+1\right)\left(\vec{\Sigma}\,\vec{V}\right)=
-\left(\vec{\Sigma}\,\vec{V}\right)\left(\vec{\Sigma}\,\vec{l}+1\right)~.\]

Remembering $\beta\vec{\Sigma}=\vec{\Sigma}\beta$, this result can
be carry over the Dirac's $K$ operator
\[K\left(\vec{\Sigma}\,\vec{V}\right)=-\left(\vec{\Sigma}\,\vec{V}\right)K~.\]
It is evident, that the class of anticommuting with $K$(or $K$-
odd) operators is not confined by these operators only - any
operator of type $\hat{O}\left(\vec{\Sigma}\,\vec{V}\right)$,
where $\hat{O}$ is commuting with $K$, but otherwise arbitrary,
also is $K$-odd.

Let mention, that in the framework of constraints of above
theorem, the following very useful relation takes place
\begin{equation}\label{usrel}K\left(\vec{\Sigma\vec{V}}\right)=
-\beta\left(\vec{\Sigma}~,~{1\over
2}\left[\vec{V}\times\vec{l}-\vec{l}\times\vec{V}\right]\right)~.
\end{equation}
We see, that the antisymmetrized vector product, familiar for LRL
vector, appears on the right- hand-side of this relation.

Important special cases, resulting from the above theorem, include
$\vec{V}=\hat{\vec{r}}$ (unit radial vector), $\vec{V}=\vec{p}$
(linear momentum) and $\vec{V}=\vec{A}$ (LRL vector), which has
the following form \cite{TT}

\begin{equation}\vec{A}=\hat{\vec{r}}-{i\over 2ma}\left[\vec{p}\times\vec{l}-\vec{l}\times\vec{p}\right]~.\end{equation}

According to (\ref{usrel}), there appears one relation between
these three odd operators

\begin{equation}\vec{\Sigma}\vec{A}=\vec{\Sigma}\hat{\vec{r}}+{i\over ma}\beta K\left(\vec{\Sigma}\vec{p}\right)~.\end{equation}
So far as $[\beta~,~K]=0$, it follows that $\left\{K~,~\beta
K\left(\vec{\Sigma}\vec{p}\right)\right\}=0$ and
$K\left(\vec{\Sigma}\vec{p}\right)$ can be used as a permissible
$K$-odd operator.

Our purpose is to construct such combination of $K$-odd operators,
which would be commuting with Dirac Hamiltonian. We can solve this
task by step by step. As a first trial expression let consider the
following operator

\[A_1=x_1\left(\vec{\Sigma}\hat{\vec{r}}\right)+ix_2K\left(\vec{\Sigma}\vec{p}\right)~.\]
Here the coefficients are chosen in such a way, that $A_1$ be
Hermitian, when $x_1~,~x_2$ are arbitrary real numbers. These
numbers must be determined from the requirement of commuting with
$H$. Let calculate

\[[A_1,H]=x_1\left[\left(\vec{\Sigma}\hat{\vec{r}}\right),H\right]
+ix_2\left[K\left(\vec{\Sigma}\vec{p}\right),H\right]~.\]
Appearing here commutators can be calculated easily. The result is
\[[A_1,H]=x_1{2i\over r}\beta K\gamma^5 - x_2{a\over
r^2}K\left(\vec{\Sigma}\hat{\vec{r}}\right)~.\] One can see, that
the first term in right-hand side is antidiagonal, while the
second term is diagonal. So this expression never becomes
vanishing for ordinary real numbers $x_1~,~x_2$. Therefore we must
perform the second step: one has to include new odd structure,
which appeared on the right-hand  side of above expression. Hence,
we are faced to the new trial operator
\begin{equation}A_2=x_1\left(\vec{\Sigma}\hat{\vec{r}}\right)+ix_2K\left(\vec{\Sigma}\vec{p}\right)+ix_3K\gamma^5f(r)~.\end{equation}
Here $f(r)$ is an arbitrary scalar function to be determined. Let
calculate new commutator. We have \[[A_2,H]=x_1{2i\over r}\beta
K\gamma^5 - x_2{a\over r^2}K\left(\vec{\Sigma}\hat{\vec{r}}\right)
-x_3f'(r)K\left(\vec{\Sigma}\hat{\vec{r}}\right)-ix_32m\beta
K\gamma^5f(r)~.\] Grouping diagonal and antidiagonal matrices
separately and equating this expression to zero, we obtain
equation \[K\left(\vec{\Sigma}\hat{\vec{r}}\right)\left({a\over
r^2}x_2+x_3f'(r)\right) +2i\beta K\gamma^5\left({1\over
r}x_1-mx_3f(r)\right)=0~.\] This equation is to be satisfied, if
diagonal and antidiagonal terms become zero separately, i.e.,
\[{a\over r^2}x_2=-x_3f'(r)~~~~,~~~~{1\over
r}x_1=mx_3f(r)~.\] First of all, let us integrate the second
equation in the interval $(r~,~\infty)$. It follows
\[x_3f(r)=-{a\over r}x_2~.\]
Accounting this in the first equation, we have
\[x_2=-{1\over ma}x_1~.\]
Then we obtain also
\[x_3f(r)=-{1\over mr}x_1~.\]
Therefore finally we have derived the following operator  which
commutes with Dirac Hamiltonian
\begin{equation}\label{opcomDH}A_2=x_1\left\{\left(\vec{\Sigma}\hat{\vec{r}}\right)-{i\over ma}K\left(\vec{\Sigma}\vec{p}\right)+{i\over m}K\gamma^5{1\over r}\right\}~.\end{equation}
It is $K$-odd, in accord with above theorem.

If we turn to usual $\vec{\alpha}$ matrices using the relation
$\vec{\Sigma}=\gamma^5\vec{\alpha}$ and taking into account the
expression (\ref{Dirham}) of the Dirac Hamiltonian, $A_2$ can be
reduced to the more familiar form ($x_1$, as unessential common
factor, may be dropped)
\[A_2=\gamma^5\left\{\vec{\alpha}\hat{\vec{r}}-{i\over
ma}K\gamma^5\left(H-\beta m\right)\right\}~.\] This expression is
nothing but the Johnson-Lippmann operator, (\ref{JLoper}).

In order to clear up its physical meaning, remark, that equation
(\ref{opcomDH}) may be rewrite in the form
\[A_2=\vec{\Sigma}\left(\hat{\vec{r}}-{i\over
2ma}\beta\left[\vec{p}\times\vec{l}-\vec{l}\times\vec{p}\right]\right)+{i\over
mr}K\gamma^5~.\] It is evident, that in nonrelativistic limit,
when $\beta\rightarrow 1$ and $\gamma^5\rightarrow 0$, this
operator coincides to the projection of LRL vector on the electron
spin direction, $A_2=\vec{\Sigma}\vec{A}$ or because of
$\vec{l}\vec{A}=0$ it is a projection on the total $\vec{J}$
momentum direction.

After this it is clear that the Witten algebra can be derived by
identifying supercharges as
\[Q_1=A~,~~~~Q_2=i{AK\over k}~.\]
It follows that
\[\{Q_1,Q_2\}=0~,~~\mbox{and}~~Q_1^2=Q_2^2=A^2~.\]
This last factor may be identified as a Witten Hamiltonian (N=2
supersymmetry).

As for spectrum, it is easy to show that the following relation
holds \cite{KA}
\[A^2=1+\left({K\over a}\right)^2\left({H^2\over m^2}-1\right)~.\]
Because all operators entered here commute with $H$, we can
replace them by their eigenvalues. Therefore, we obtain energy
spectrum pure algebraically. In this respect it is worth-while to
note full analogy with classical mechanics, where closed orbits
were derived by calculating the square of the LRL vector without
solving the differential equation of motion, \cite{Gol}. In
conclusion we are convinced that the degeneracy of spectrum
relative to interchange $k\rightarrow -k$ is associated to the
existence of conserved J.-L. operator which takes its origin from
the LRL vector. It is also remarkable, that the same symmetry is
responsible for absence of the Lamb shift in this problem.
Inclusion the Lamb-shift terms into the Dirac Hamiltonian breaks
commutativity of $A$ with $H$.

In conclusion we should like to  express  our cincere appreciation
to professors Misha Zviadadze, Gela Devidze and Badri Magradze for
their critical remarks.

This work was supported in part by the reintegration Grant No.
FEL. REG. 980767.

\end{document}